\documentclass[10pt,aps,prb,showpacs,linenumbers,superscriptaddress,twocolumn,citeautoscript]{revtex4-2}

\usepackage{graphicx}  
\usepackage{dcolumn}   
\usepackage{bm}        
\usepackage{amssymb}  
\usepackage{amsmath}
\usepackage{hyperref,color,braket}
\usepackage{textcomp} 

\usepackage{calrsfs} 
\usepackage{mathrsfs} 

\newcommand{\KeyWords}[1]{\vspace{2mm}\par\noindent{\small{\em Keywords\/}: #1}}

\mathchardef\mhyphen="2D

\DeclareUnicodeCharacter{0190}{\ensuremath{\epsilon}}
\begin{document}
\title{Near-Field Mechanical Fingerprints for THz Sensing of 'Hidden' Nanoparticles in Complex Media}

\author{Ricardo Mart\'{\i}n Abraham-Ekeroth}
\email{mabraham@ifas.exa.unicen.edu.ar}
\affiliation{Instituto de F\'{\i}sica Arroyo Seco, Universidad Nacional del Centro de la Provincia de Buenos Aires, Pinto 399, 7000 Tandil, Argentina}
\affiliation{Centro de Investigaciones en F\'{\i}sica e Ingenier\'{\i}a del Centro de la Provincia de Buenos Aires (UNCPBA-CICPBA-CONICET), Campus Universitario, 7000 Tandil, Argentina.}
\affiliation{Grup de Recerca d'\`{O}ptica de Castell\'{o} (GROC), Institut de Noves Tecnologies de la Imatge (INIT), Universitat Jaume I (UJI), Castell\'{o} de la Plana, 12071, Spain.}
\author{Dani Torrent}
\affiliation{Grup de Recerca d'\`{O}ptica de Castell\'{o} (GROC), Institut de Noves Tecnologies de la Imatge (INIT), Universitat Jaume I (UJI), Castell\'{o} de la Plana, 12071, Spain.}
\date{\today}
\begin{abstract}
Terahertz (THz) spectroscopy holds transformative potential for non-invasive sensing, yet characterizing individual nanoparticles in complex biological environments remains challenging due to the far-field diffraction limit. While near-field dipolar theory is well established, its application to characterizing/identifying nanoparticles immersed in complex media at THz frequencies is largely unexplored.

This work utilizes numerical simulations of magneto-optical (MO) heterodimers -comprising $n$-doped Indium Antimonide ($n\mhyphen InSb$) and isotropic or birefringent particles (e.g., $SiO_{2}$, $GaSe$)- under counter-propagating, circularly polarized THz illumination. We demonstrate that while far-field observables like absorption cross-sections are often dominated by the MO-active particle, mechanical variables—specifically induced binding forces and spin/orbital torques—exhibit superior sensitivity for detecting "hidden" neighboring components. Because these mechanical signatures depend directly on near-field interactions, they provide higher information density regarding interparticle coupling.

Key findings reveal material-specific spectral "hotspots" and "zeros" that serve as robust calibration markers even within dispersive biological surrogates. We show that the spin torque on non-MO particles is significantly modified by MO-neighbor proximity, a phenomenon controllable via static magnetic fields. Furthermore, these variables exhibit high angular sensitivity in perpendicular configurations. Our results provide a roadmap for using optomechanical signatures as high-resolution detectors for \textit{in-vivo} diagnostics, signal transduction, and low-energy nanocircuit control.
\end{abstract}
\nolinenumbers
\maketitle


\section{Introduction}

The exploration of terahertz (THz) spectra has transitioned from a significant technological "gap" into one of the most promising frontiers for non-invasive sensing and theragnostic applications \cite{leitenstorfer_2023_2023}. THz radiation possesses unique physical properties that facilitate deep penetration through non-polar media -including polymers, ceramics, and dry biological tissues- which are often diffusive or opaque at visible or infrared wavelengths. Simultaneously, this regime is characterized by an exquisite sensitivity to the hydration states and rotational-vibrational transitions of polar environments, enabling real-time monitoring of molecular dynamics and phase changes at high resolution \cite{penkov_terahertz_2023}. THz radiation's non-ionizing nature and sensitivity to tissue structure make it ideal for intraoperative mapping and cancer diagnosis \cite{duan_application_2025}. Higher water content, combined with structural variations in tumoral tissues (increased cell and protein density), results in higher THz absorption and modified refractive indices. The "hyperthermia effect" offers another therapeutic path: nanoparticles (NPs) can be irradiated with near-infrared or THz beams to induce surface plasmon excitation, raising the temperature of water in local cancer cells. This temperature change can then be imaged in real-time with THz radiation \cite{naccache_terahertz_2017}. Recent advancements have introduced novel photoactive materials, optimized device structures, and sophisticated detection mechanisms such as THz Time-Domain Spectroscopy (THz-TDS) and scanning near-field optical microscopy (THz-SNOM) \cite{gezimati_terahertz_2023,yan_terahertz_2025-1}. These developments have paved the way for advanced biomedical research, security screening, and 6G communications. 

Central to these advancements is the interaction between THz radiation and nanoscale matter, where the wavelength-to-particle size ratio allows for the manipulation of NPs as local perturbations of an effective medium rather than mere scattering centers. THz radiation overlaps with the energy range of water relaxational motions and the intermolecular vibrations of biomacromolecules, making it a unique probe for the hydration shells that surround proteins, DNA, and phospholipids \cite{fan_use_2019}. These hydration shells are not static; they exhibit molecular dynamics on timescales ranging from picoseconds to microseconds, influencing the functional behavior of biological systems \cite{penkov_terahertz_2023}. At the nanoscale, the physical response of objects to THz radiation is predominantly characterized by the Rayleigh scattering regime \cite{huang_enhancement_2015}. With wavelengths spanning approximately to in the $1$-$50$ THz range, a nanoparticle system remains deep within this limit. In this framework, the particles do not independently diffuse the incident beam but function as local hotspots or perturbations that can maintain the coherence of the electromagnetic (EM) field. This coherence is fundamental for optomechanical manipulation, allowing for the precise application of optical forces and torques through interference and phase control \cite{roichman_optical_2008}. 

In particular, a significant advancement in THz photonics involves the use of Indium Antimonide ($InSb$) and its most notable magneto-optical (MO) counterpart, $n$-doped $InSb$ (briefly $n\mhyphen InSb$) \cite{choi_metallic_1991,chochol_magneto-optical_2016-1,sadrara_electric_2019-1,charca-benavente_high-q_2025}. Unlike many materials that require high-intensity magnetic fields to exhibit MO effects, $n\mhyphen InSb$ demonstrates strong responses at modest fields due to its exceptionally low effective carrier mass \cite{chochol_magneto-optical_2016-1,gabbani_magnetoplasmonics_2022}. This property enables the excitation of surface magnetoplasmons (SMPs) at room temperature, facilitating the creation of non-reciprocal devices that can steer or guide THz radiation at subwavelength scales \cite{gerasimov_exploiting_2020}. 
  
In parallel with plasmonic materials, the use of $III$-$VI$ semiconductors like Gallium Selenide ($GaSe$) provides a different modality for THz interaction. $GaSe$ is a layered, negative uniaxial crystal \cite{fletcher_measurement_2017,yu_terahertz_2005} known for its high transparency and sharp phonon resonances in the THz regime. $GaSe$'s large birefringence and high figure-of-merit make it an ideal candidate for phase shifters, THz generation through optical rectification, and difference-frequency generation (DFG). The material's mechanical response is orientation-dependent, which, when integrated into a nanoparticle system, allows for the detection of individual particles based on their specific crystallographic alignment and phonon signatures.   

The current landscape of THz sensing is dominated by two primary methodologies: far-field (FF) spectroscopy and near-field (NF) imaging \cite{penkov_terahertz_2023}. While FF systems, such as THz-TDS, are effective for identifying collective molecular signatures and macroscopic hydration states, they are inherently limited by the diffraction limit. This constraint hinders the ability of FF systems to distinguish between individual NPs within a complex cluster, as the scattering and absorption cross-sections often overlap and obscure local details. To overcome the diffraction limit, researchers have turned to THz-SNOM \cite{yan_terahertz_2025-1}. This technique utilizes a subwavelength aperture or a sharp metal nanotip to confine the EM field into a "hotspot" far smaller than the wavelength of the radiation \cite{huang_enhancement_2015}. Recent reviews highlight the emergence of four distinct THz-SNOM modalities, offering spatial resolutions as fine as $20$ nm \cite{klarskov_nanoscale_2017}. 
   
The integration of metamaterials has further enhanced the sensitivity of THz biosensors \cite{wang_recent_2025}. By engineering metallic or dielectric micro-structures, researchers can create split-ring resonators (SRRs) and chiral surfaces that exhibit unique EM resonance phenomena such as Fano resonances and toroidal modes \cite{zhang_terahertz_2023}. These structures are highly sensitive to the local refractive index, enabling the detection of trace biochemical substances, viral pathogens, and glioma cell subtypes \cite{wang_recent_2025}. For instance, sensitivity can be improved by a factor of $13$ by reducing the gap width in SRRs from $3 \mu$m to $200$ nm, facilitating the detection of viruses with diameters as small as $30$ nm \cite{park_sensing_2017}.

Despite the successes of NF imaging, a fundamental bottleneck remains in the ability to characterize and manipulate hidden NPs within complex, gelatinous environments typical of biological media. Recent theoretical work proposes that MO binding forces and torques can serve as high-resolution "local probes" that provide information inaccessible to FF observables \cite{bahk_detection_2022,abraham-ekeroth_numerical_2023}. Optical binding refers to the inter-particle interaction induced by light between two or more objects \cite{yan_guiding_2013,forbes_optical_2020-1}. In the Rayleigh regime, this interaction can be modeled using a coupled-dipole approach \cite{novotny_principles_2006,edelstein_magneto-optical_2021-1}. When two particles are illuminated by counter-propagating plane waves, they experience forces that depend on the phase and intensity gradients of the local field.   
The introduction of a static magnetic field to an MO-active dimer (e.g., $n\mhyphen InSb$) breaks the symmetry of the system. This non-reciprocity results in tunable inter-particle forces and spins that can be used to assembly or disassemble "optical matter" at will \cite{abraham-ekeroth_modeling_2025}. 
Crucially, while FF absorption may be dominated by a "large" MO particle, other NF mechanical magnitudes remain highly sensitive to the proximity and material properties of a second, maybe smaller particle. In addition to linear forces, the angular momentum of the THz field -both spin (SAM) and orbital (OAM)- can be transferred to the particles, exerting torques \cite{yevick_photokinetic_2017,edelstein_magneto-optical_2019}. Spin torque arises from the interaction between the field's polarization and the particle's dipole moment, while orbital torque is related to the spatial distribution of the field \cite{nieto-vesperinas_optical_2015-1}. In MO systems, these torques are coupled and can be modified by the external magnetic field, providing a unique mechanical signature that identifies the type of mode being excited \cite{abraham-ekeroth_thz_2025-1}. In addition, MO nanocarriers could be loaded with therapeutic molecules and guided within the body using external magnetic fields, while their position is monitored through the mechanical fingerprints described above. Single NPs, such as gold NPs or $InSb$-based resonators, can be used to "turn off" or modulate THz resonances in slot antennas, providing a way to detect ultra-low densities of biomolecules and lipid vesicles  \cite{eliahoo_viscoelasticity_2024}.    

Following the methodology established in previous work -utilizing a dual counter-propagating plane wave setup to achieve a specific stationary field \cite{abraham-ekeroth_numerical_2023,abraham-ekeroth_thz_2025-1}- the response of MO heterodimers (i.e., dimers of dissimilar particles) is analyzed in both parallel and perpendicular configurations relative to an external magnetic field. This dimer arrangement serves as a fundamental geometric model to illustrate the underlying physical concepts of interparticle coupling. It is demonstrated that while traditional FF observables, such as absorption cross-sections, often fail to distinguish individual particles within a dimer, NF mechanical variables offer superior sensitivity. Specifically, it is shown that induced binding forces and spin/orbital torques provide a robust method for identifying "hidden" particles and characterizing their resonances, even when they are obscured by the dominant absorption profile of the MO-active object or bigger particle. These results pave the way for advancing THz-based theragnostics, provide novel designs of THz antennas, and the development of nanorobots/ nanomachines working at low energy. 

\section{Methodology}

\begin{figure}
	\centering
	\includegraphics[width=8.5cm,keepaspectratio]{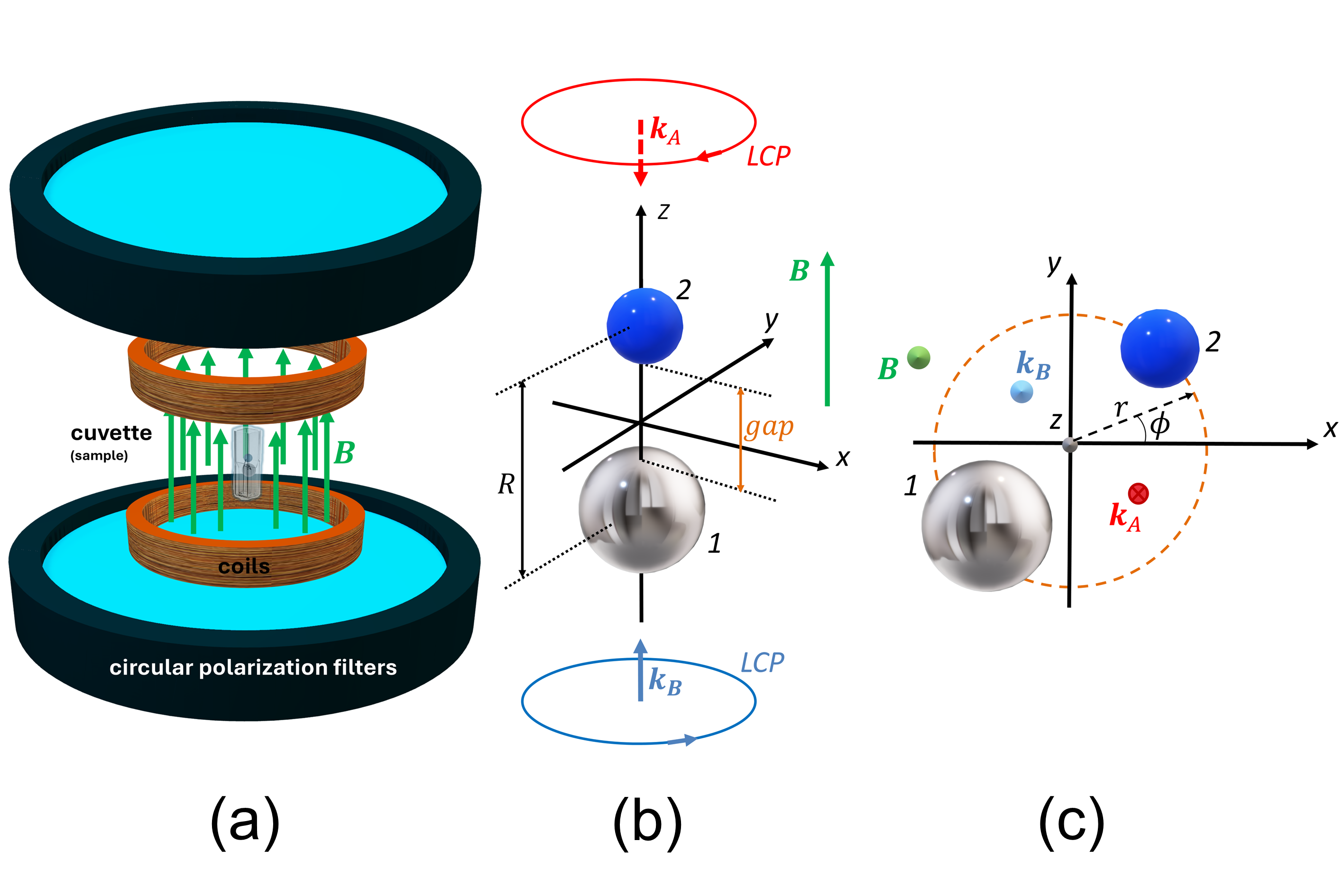}
	\caption{\label{fig:1_Config} (Color online) (a) Schematic of a possible thought setup to study magneto-optical samples consisting of dimers with reduced mobility. The dimers are contained in a cuvette filled with, e.g., viscoelastic gel. The external (static) magnetic field ($\vec{B}$) is generated by Helmholtz coils, and the illumination field at THz is achieved by using appropriate polarization filters. Such beam is built with two counter-propagating plane waves, with wavevectors $k_A$ and $k_B$, both waves possessing left circular polarization (LCP). (b-c) Details of the geometry and composition of the configurations studied in this work; the illuminating beam and $\vec{B}$ are always parallel to the $z$-axis of the cartesian coordinate system. (a) Parallel configuration; or dimer parallel to $z$. (b) Perpendicular configuration, where the dimer is held along the $xy$ plane at $z=0$. The first particle (silvered, $\# 1$) is made of n-doped $InSb$; the other particle (blue, $\#2$) is made of an isotropic or birefrigent material.}
\end{figure}

To explore the behavior of NF mechanical variables, a theoretical framework is implemented based on the configuration illustrated in Fig. \ref{fig:1_Config}a. This setup serves as a proof-of-concept, simulating a dual counter-propagating plane wave arrangement with identical left-handed circular polarization (LCP). Such a configuration is idealized to ensure that, in a reciprocal system, the net radiation pressure vanishes \cite{abraham-ekeroth_numerical_2023}; this allow inter-particle forces to be isolated and studied with high precision -a critical requirement for future experimental realization. The particles are modeled as being immersed in a cuvette filled with a viscoelastic gel, such as Polyethylene (PE) or Paraffin Wax \cite{miranda_properties_2021}. In this simulated environment, the gel serves to damp thermal fluctuations and mimic the viscoelastic response of biological tissues, effectively isolating the light-induced degrees of freedom \cite{eliahoo_viscoelasticity_2024}. The embedding media are assigned representative THz permittivity values, ranging from $\epsilon_{rb} = 1.75$ (PBS) to $\epsilon_{rb} \approx 4.0$ (hydrogels), to ensure the results remain relevant to a broad class of biological surrogates \cite{gezimati_terahertz_2023}. Within this framework, the dimer's MO response is activated by a uniform magnetic field ($B$), conceptually generated by Helmholtz coils (Fig. \ref{fig:1_Config}a). To provide a comprehensive analysis of the system's mechanical coupling, two fundamental geometric configurations are enforced: parallel (aligned with the $z$-axis, Fig. \ref{fig:1_Config}b) and perpendicular (lying in the $xy$-plane, Fig. \ref{fig:1_Config}c). While these configurations are treated here as a computational model, i.e., the ones in Figs.~\ref{fig:1_Config}b-c, the sketch in Fig.~\ref{fig:1_Config}a is designed to define the parameters and expected observables for subsequent THz-trap experimental validations.
As the particles considered are much smaller than the working wavelengths, they can be roughly modeled by their dipolar response. The particles $\#1$ and $\#2$ are then represented by dipole moments $\vec p_{1}$-$\vec p_{2}$ respectively, defined as 
\begin{align} 
	\vec p_{1} = \epsilon_0 \epsilon_{rb} \hat \alpha_1 \vec E_{eff,1}, \label{eq-dipole-p1} \\ 
	\vec p_{2} = \epsilon_0 \epsilon_{rb} \hat \alpha_2 \vec E_{eff,2} \label{eq-dipole-p2}
\end{align}
where we use the definitions of the "effective" fields, $\vec E_{eff,j}$, which are coupled by the Green tensor $\hat G$ \cite{abraham-ekeroth_numerical_2023} as
\begin{align}  
	\vec E_{eff,1} = \vec E_{0,1} + k^2_b \hat G \hat \alpha_2 \vec E_{eff,2}, \label{eq-Eeff1} \\ 
	\vec E_{eff,2} = \vec E_{0,2} + k^2_b \hat G \hat \alpha_1 \vec E_{eff,1}. \label{eq-Eeff2}
\end{align}
where $k_b = n_{rb} k_0$, such that $n_{rb}$ is the refractive index and $k_0$ the vacuum wavenumber. We have also used the fact that the Green tensor is symmetric, i.e., $\hat G_{12} = \hat G_{21} \equiv \hat G$. The explicit definition of $\hat G$ can be found in Ref.~\cite{novotny_principles_2006}. Here, $\epsilon_0$ is the vacuum permittivity, $\vec E_{0,n}$ is the incident electric field at the dipoles' positions $\vec{r}_n$, where $n=\{1,2\}$, and $\hat{\alpha}_n$ is the polarizability tensor that represents the particles, which includes the radiative corrections when it is expressed as follows \cite{albaladejo_radiative_2010}:
\begin{equation}
	\label{eq-alpha}
	\hat \alpha = \left(\hat \alpha_{n,0}^{-1} - \frac{ik^3_b\hat I}{6\pi}\right)^{-1}
\end{equation}
where $\hat \alpha_{n,0}$ are the quasistatic polarizabilities, which for spherical dipolar particles can be given by
\begin{equation}
	\label{eq-alpha0}
	\hat \alpha_{n,0}^{-1} = \frac{1}{V_n} \left(\hat I/3 + [\hat \epsilon_{r,n} - \hat I]^{-1} 
	\right).
\end{equation}
where $V_n$ are the particle volumes and $\hat \epsilon_{r,n}$ their relative dielectric tensors. The explicit definition of the MO permittivity tensor function for $n\mhyphen InSb$ and the spectral behaviour of their components can be found in Ref.~\cite{abraham-ekeroth_numerical_2023}. 
It is not difficult to show that, by using Eqs.~\ref{eq-Eeff1}-\ref{eq-Eeff2} into Eq.~\ref{eq-dipole-p1}, leads to the solution
\begin{align} 
	\label{eq-p1-solved} 
	\vec p_{1} = \epsilon_0 \epsilon_{rb}\hat F \left(\vec E_{0,1} + k^2_b \hat G \hat \alpha_2 \vec E_{0,2}\right),
\end{align}
where $\hat F = \left(\hat{\alpha_1}^{-1} - k^4_b \hat G \hat \alpha_2 \hat G \right)^{-1}$. Inserting this result into Eq.~\ref{eq-Eeff2} by means of def.~\ref{eq-dipole-p1}, we get
\begin{align} 
\label{eq-p2-solved} 
\vec p_{2} = \epsilon_0 \epsilon_{rb} \hat \alpha_2 \left(\vec E_{0,2} + \frac{k^2_b}{\epsilon_0 \epsilon_{rb}} \hat G \vec p_{1} \right). 
\end{align}
The illumination consists of a superposition of two counter-propagating, left-handed circularly polarized (LCP) plane waves with the same intensity $I_0$   \cite{cameron_optical_2014,edelstein_magneto-optical_2019, li_advances_2022}, see Fig.~\ref{fig:1_Config}, namely,
\begin{equation}
	\label{eq-E0}
	\vec E_{0} = \frac{E_{0}}{\sqrt{2}}\left[\left(\check{x}+i\check{y}\right)e^{ik_bz}+\left(\check{x}-i\check{y}\right)e^{-ik_bz}\right].
\end{equation}
It is worth mentioning why this standing-wave field is chosen. This illumination easily creates a total field with zero-average values of $\nabla |\vec{E}|^2$, $\vec{S}$, and $\nabla \times \vec{J}_{spin}$, where $\vec{E}$, $\vec{S}$, and $\vec{J}_{spin}$ are the electric, Poynting, and spin density field. In other words, this total field induces zero net force on an isotropic particle in the absence of a static magnetic field $\vec{B}$,  \cite{edelstein_magneto-optical_2019}. Its total spin is always equal to zero, implying a null gradient force, while its helicity has a constant positive value, which implies a constant extinction force when $\vec{B}$ is on \cite{abraham-ekeroth_numerical_2023}. Besides avoiding complex optical traps, this illumination is homogeneous across space, which could help bind multiple particles to form large-scale optical matter, for example, using NPs and laser illumination. 

The absorption cross section of the system can be calculated once the dipole moments are known by
\begin{align}
	& \sigma_{abs}=\frac{k_b}{\epsilon_0^2\epsilon_{rb}^2|\vec{E}_0|^2}Im\left\{\vec p_1\cdot\left(\hat \alpha_{10}^{-1}\vec p_1\right)^* + \vec p_2\cdot\left(\hat \alpha_{20}^{-1}\vec p_2 \right)^*\right\}	
\end{align}
The $i$-component of the forces exerted on each particle can be obtained from the time-averaged force within the Rayleigh approximation \cite{chaumet_time-averaged_2000}. This is 
\begin{eqnarray}
	\label{eq-FcDDA}
	F_{1,i}& = &\frac{1}{2}Re\{\vec p^{\, t}_1[\partial_i\vec{E}^{*}(\vec{r},\omega)|_{\vec{r}=\vec{r}_1}\} \\
	F_{2,i}& = &\frac{1}{2}Re\{\vec p^{\, t}_2[\partial_i\vec{E}^{*}(\vec{r},\omega)|_{\vec{r}=\vec{r}_2}\}
\end{eqnarray}
where the derivatives of the total field $\partial_i\vec{E}(\vec r,\omega)|_{\vec r=\vec r_n}$ at the dipoles' positions $\vec{r}_n$ can be obtained from \cite{chaumet_coupled_2007}:
\begin{align}
	\label{eq-derivsEr}
	& \partial_i\vec{E}(\vec{r},\omega)|_{\vec{r}=\vec{r}_1}=\partial_i\vec{E}_0(\vec{r},\omega)|_{\vec{r}=\vec{r}_1}+ \nonumber \\ 
	& +\frac{k^2_b}{\epsilon_0}(\partial_i\hat G(\vec{r},\vec{r}_2))_{\vec{r}=\vec{r}_1}\vec{p}_2]\}  \\
	&\partial_i\vec{E}(\vec{r},\omega)|_{\vec{r}=\vec{r}_2}=\partial_i\vec{E}_0(\vec{r},\omega)|_{\vec{r}=\vec{r}_2} + \nonumber \\ 
	& +\frac{k^2_b}{\epsilon_0}(\partial_i\hat G(\vec{r}_1,\vec{r}))_{\vec{r}=\vec{r}_2}\vec{p}_1]\} 
\end{align}
The total force exerted on the dimer results from adding the force components for each particle, namely, $F_{tot,i} = F_{1,i}+F_{2,i}$. In particular, the net radiation pressure for the dimer under the illumination given by Eq.~\ref{eq-E0} is defined by taking $i=3$, or the $z$ components, as
\begin{equation}
	\label{eq-FzSys}
	F_{tot,z} = F_{1,z}+F_{2,z}
\end{equation}
Another useful mechanical variable is the binding force, which in the present case is defined as
\begin{equation}
	\label{eq-BindingF}
	\Delta = \left(\vec{F}_1 - \vec{F}_2\right)\cdot \check{n}
\end{equation}
where $\check{n}=\frac{\vec{r}_2-\vec{r}_1}{|\vec r_2-\vec r_1|}$ is the dimer's versor. The optical torques can also be calculated, as given in Ref.~\cite{chaumet_electromagnetic_2009}:
\begin{align}
	\label{eq-Tqs1}
	& \vec N_{spin,1} = \frac{1}{2\epsilon_0}Re\left\{\vec p_{1} \times \left[\left(\hat \alpha_0^{-1}\right)^*\vec p_1^{\, *}\right]\right\} \\
	& \vec N_{orb,1} = \vec r_{1} \times \vec F_{1} \\
	& \vec N_{1} = \vec N_{spin,1} + \vec N_{orb,1}
\end{align}
\begin{align}
	\label{eq-Tqs2}
	& \vec N_{spin,2} = \frac{1}{2\epsilon_0}Re\left\{\vec p_{2} \times \left[\left(\hat \alpha_0^{-1}\right)^*\vec p_2^{\, *}\right]\right\} \\
	& \vec N_{orb,2} = \vec r_{2} \times \vec F_{2} \\
	& \vec N_{2} = \vec N_{spin,2} + \vec N_{orb,2}
\end{align}
The definitions of the orbital and spin torques were discussed previously in Refs.~\cite{nieto-vesperinas_optical_2015-1,chaumet_electromagnetic_2009}, among others. The spin torques are always defined with respect to the centers of the particles. Otherwise, the reference system is located at the dimer's center of mass, and orbital torques are set. 

\section{Results and Discussion}

Specifically, all dimers calculated are composed of a $n\mhyphen InSb$ dipole as a MO active particle ($\# 1$) of radius $50$ nm and an another non-MO, smaller particle (particle $\# 2$, radius $25$ nm). The inter-particle gap in the dimers is kept relatively small, i.e., $gap=10$ nm, since the interest lies in the NF sensing. Consequently, all the dimers are small compared to the wavelength $c/frequency$ of the illumination, their scattering cross-sections being negligible and the FF spectra are thus entirely characterized by the absorption cross-sections. It is also necessary to point out that, under the illumination given by Eq.~\ref{eq-E0}, the radiation pressures become negligible for all the systems studied in this work. No average $z$-forces practically exist in the studied systems, i.e., $F_{tot,z} \sim 0$ in Eq.~\ref{eq-FzSys}. 

\subsection{Single Particle}

\begin{figure*}
	\centering
		\includegraphics[width=15cm,keepaspectratio]{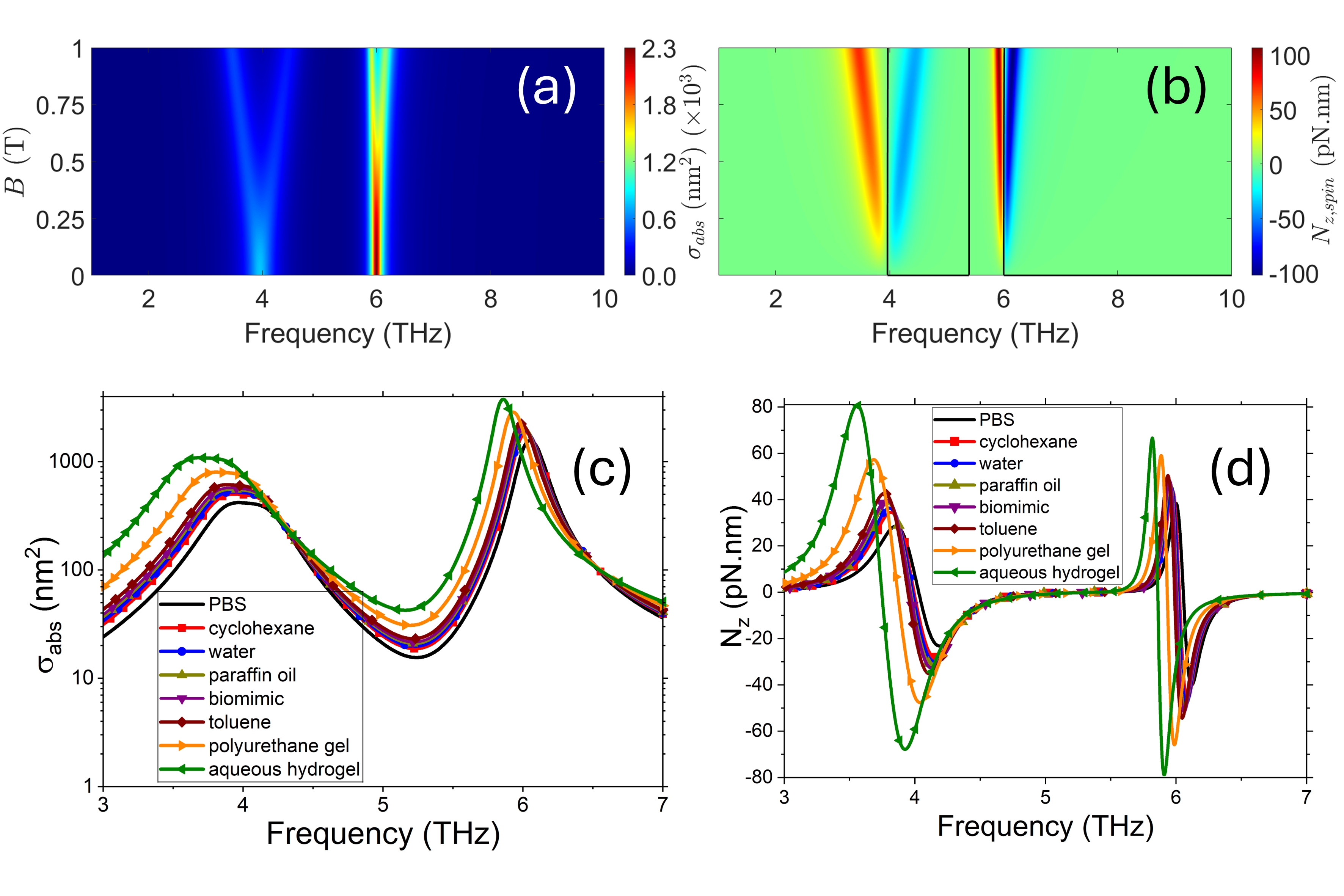}
	\caption{\label{fig:2_SingleParticle} (Color online) Response by a single, $100$ nm, MO particle ($n\mhyphen InSb$) under the illumination described in Eq.~\ref{eq-E0}. (a) Map of absorption cross-section's spectra as a function of the external (static) magnetic field. (b) idem -a- for the spin torque exerted on the particle. The embedding medium is "biomimic". Black line highlights the zeros calculated for this observable. (c) Spectra showing the cross-section's  resonance peaks for several embedding media. Here, $B= 0.25$ T. (d) Idem -c- for the resonant features in the spin torque.}
\end{figure*}

 The baseline response of a single $100$ nm $n\mhyphen InSb$ particle reveals the fundamental MO resonances that will appear below as well for dimers. As a matter of fact, two maps in Figs.~\ref{fig:2_SingleParticle}a-b show the single-particle absorption and spin torques when the frequency and $B = \left|\vec B\right|$ are swept. Specifically, the absorption's map (Fig.~\ref{fig:2_SingleParticle}a) shows a couple of resonances at $B=0$ ($B$ \textit{off}, bottom values) which begin to split as the value of $B$ increases (top). These correspond to phonon and plasmon polaritons typically found in $n\mhyphen InSb$ \cite{abraham_ekeroth_anisotropic_2018}. A critical observation is that the spin torque manifests these resonances in a way different than absorption does. The $z$-spin changes sign according to branches produced by the particle's resonances, and the "zeros" in the spin torque spectrum align precisely with the resonance peaks when $B$ is off. This establishes a robust method for detecting and calibrating the immersed system, as these zeros act as material-specific markers.   
In addition, a variation of the embedding medium is included in the single-particle calculations (Fig.~\ref{fig:2_SingleParticle}c-2d). Several realistic media is simulated, whose THz permittivity values correspond to: PBS, $\epsilon_{rb} = 1.75$; cyclohexane, $2.02$; water, $2.1$; paraffin oil, $2.2$; a biological average mimic (labelled \textit{biomimic} below), $2.25$; toluene, $2.38$; polyurethane gel, $3$; and aqueous hydrogel, $4$. The purpose of these results is to show that varying $\epsilon_{rb}$ between these values does not alter significantly the spectra of the single particle, with maybe the exception of high values such as those for polyurethane gel and aqueous hydrogel. Thus, the values to compare below to the dimer's are quite similar except for some slight shifts in both frequency and intensity.

\subsection{Dimer under parallel configuration}

\begin{figure*}
	\centering
		\includegraphics[width=15cm,keepaspectratio]{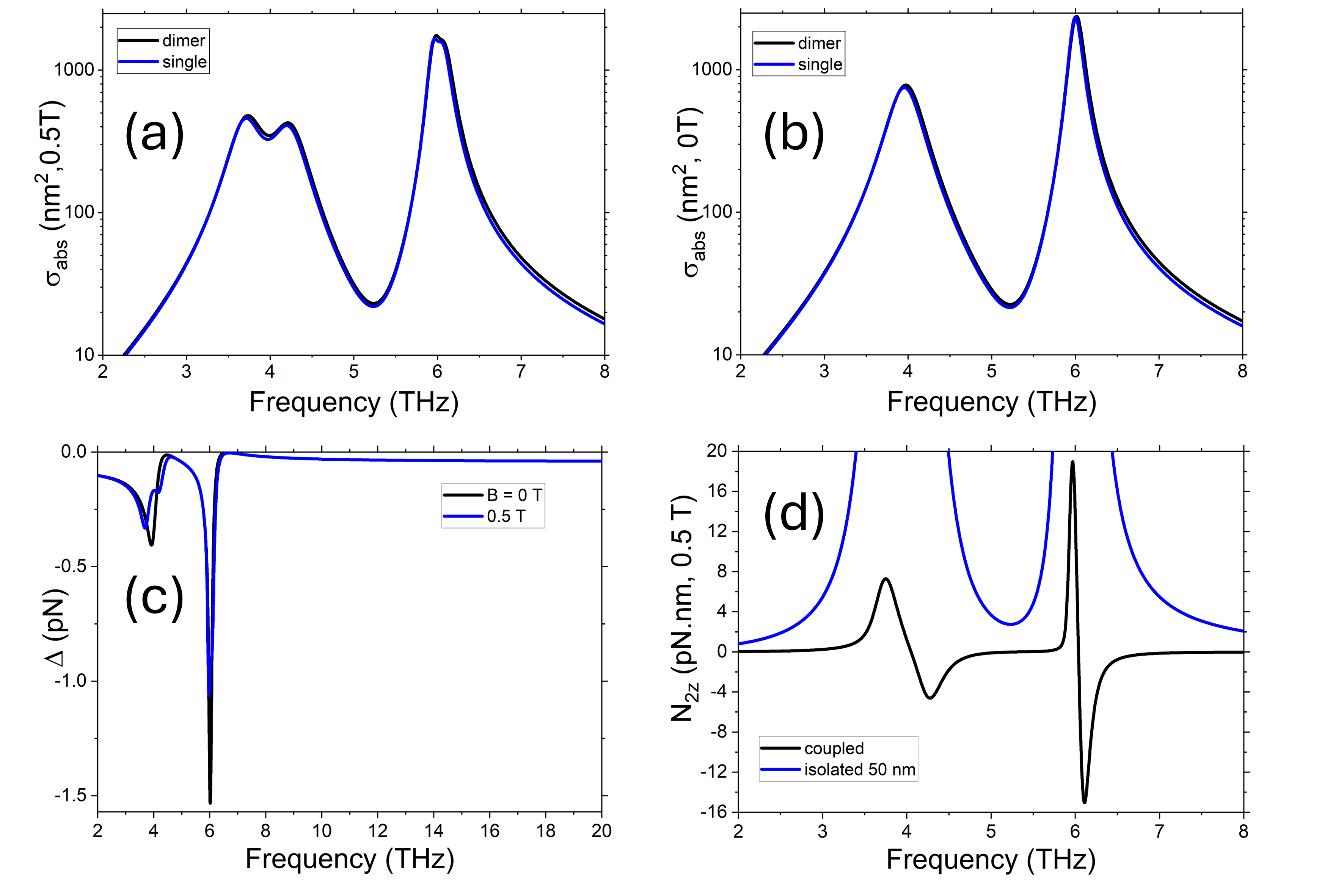}
	\caption{\label{fig:3_Spectra_parallel_conf} (Color online) Response by a MO heterodimer immersed in an average biological medium \textit{biomimic} (sizes $100$-$50$ nm, both particles made of $n\mhyphen InSb$, immersed in biomimic, $gap = 10$ nm) under parallel configuration. (a) [b] Absorption cross sections when $B = 0.5$ T [$B = 0$ T]. Black [blue] line for the dimer system [particle $\# 1$ as isolated]. (c) Binding force induced on the dimer when $B$ is on/off. (d) Spin torque exerted on particle $\# 2$; black line when it is coupled to particle $\# 1$, blue when uncoupled (as isolated).}
\end{figure*}

When a second particle is introduced in the parallel configuration, Fig.~\ref{fig:1_Config}b, the coupling produces additional degrees of freedom. In a system where both particles are $n\mhyphen InSb$, the absorption cross-sections for the dimer and an isolated particle are nearly identical, because particle $\# 1$'s absorption dominates, compare Figs~\ref{fig:3_Spectra_parallel_conf}a-b for $B$ on at $0.5$ T and off, respectively. Noticeably, both the dimer and the isolated-particle's responses (black vs. blue line) are very similar, in a way independent of the magnetic field used. Consequently, particle $\# 2$, which is smaller, results practically undetectable by the FF observable. Conversely, the NF induced mechanics -specifically binding forces and spin torques- clearly distinguish the presence of the second particle, Figs.~\ref{fig:3_Spectra_parallel_conf}c-d. The binding force in a $\left[n\mhyphen InSb, n\mhyphen InSb\right]$ dimer is typically negative, indicating repulsive states in all the resonances shown in the absorption. Note the correspondence of the resonance features in both types of spectra. In particular, a sharp dip at $6$ THz (phonon) corresponds to a strong repulsion exceeding $1$ pN when $B=0.5$ T, which reduces in magnitude if the magnetic field is further increased (Fig.~\ref{fig:3_Spectra_parallel_conf}c). 

In general, the FFs can reflect the system's absorption or scattering. However, dynamic information as movement, must in principle be connected to the NFs. NFs possess more information than FFs because of the interactions via evanescent waves. Fig.~\ref{fig:3_Spectra_parallel_conf} describes how more sensitive the NF mechanical variables can be compared to the typical FF variables. The spin torque induced on particle $\# 2$ is remarkably sensitive to the dimer's resonances, Fig.~\ref{fig:3_Spectra_parallel_conf}d. 

The spin torque induced on particle $\# 2$ is remarkably strong, black line in Fig.~\ref{fig:3_Spectra_parallel_conf}d. Noteworthy, $N_{2z}$ shows the characteristic resonances of the dimer, but in addition, the interparticle interaction drastically reduces their spectral values; compare the dimer response (black line) with the spin torque induced on particle $\# 2$ as if were isolated, blue line in Fig.~\ref{fig:3_Spectra_parallel_conf}d.
Moreover, the torque changes sign with sharp peaks each time a resonance is excited (black line), a behavior that differs significantly from an isolated particle of the same size, where the spin shows only positive peaks (blue line). This contrast underscores the role of inter-particle interaction in modifying the local mechanical response. The spin seems to detect all the broken degenerations caused by the presence of $\vec B$.

Now the study describes a heterodimer $\left[n\mhyphen InSb, GaSe\right]$ immersed in biomimic, a system which has been designed to exhibit surface phonon resonances located near the surface phonon resonances for the $n\mhyphen InSb$ particle; this is another way to explore dielectric-particle detection properties. Gallium Selenide ($GaSe$) is a hexagonal, birefringent crystal. Its permittivity tensor is diagonal when the coordinate system is aligned with its principal axes ($x, y, z$). In this orientation, the $z$-axis corresponds to the optic axis, thus the tensor is expressed as \cite{kato_sellmeier_2013}:
\begin{equation}
	\label{eq-epr-birrefringence}
	\hat \epsilon_{r2} = \begin{bmatrix} \epsilon_\perp & 0 & 0 \\ 0 & \epsilon_\perp & 0 \\ 0 & 0 & \epsilon_\parallel \end{bmatrix}
\end{equation}
where $\epsilon_\perp$ is the ordinary component and $\epsilon_\parallel$ is the extraordinary. $GaSe$ is known for being a negative uniaxial crystal, meaning that $\epsilon_\parallel < \epsilon_\perp$. It is widely used for second harmonic and THz generation, being also a van der Waals material \cite{yu_terahertz_2005,fletcher_measurement_2017}. The $\epsilon_\parallel$ component (along the optical axis) is typically much lower than the in-plane $\epsilon_\perp$ because the electronic coupling between the layers (held by weak forces) is much weaker than the covalent bonding within a single layer. This material is modeled by a Lorentz oscillator, see details in the Appendix for the effective parameters used.

\begin{figure}
	\begin{centering}
		\includegraphics[width=9cm,keepaspectratio]{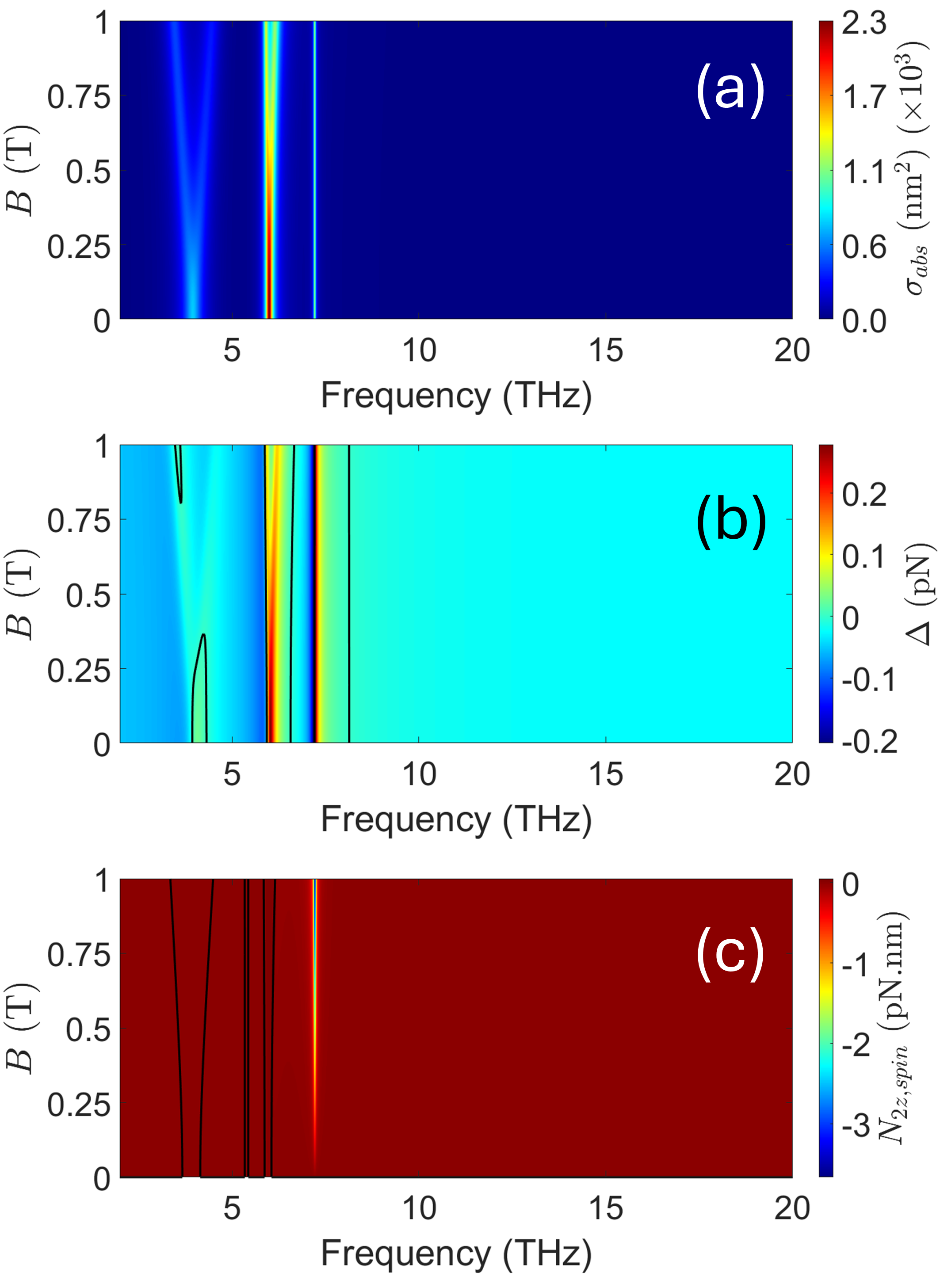}
		\par\end{centering}
	\caption{\label{fig:4_Parallel_conf_varB} (Color online) Maps of relevant observables when both the frequency and value of the external magnetic field are swept. The heterodimer consists of a $100$ nm particle ($\#1$) made of $n\mhyphen InSb$ and a $50$ nm ($\#2$) of $GaSe$ under parallel configuration, with a $10$ nm gap. (a) Absorption cross section. (b) Binding force, and (c) Spin torque exerted on the $GaSe$ particle.}
\end{figure}

For completeness, the spectra in Fig.~\ref{fig:4_Parallel_conf_varB} extend for a continuum of values of $B$ ranging from $0$ to $1$ T. Note that the absorption displays the same scheme of resonances as that for the isolated $n\mhyphen InSb$ particle at frequencies around $4$-$6$ THz, except for a single resonance at $7.2$ THz, which results independent of $B$ in the map, and is related to the  phonons excited in $GaSe$ particle ($\#2$, Fig.~\ref{fig:4_Parallel_conf_varB}a). In other words, the spectra for low frequencies ($<7$ THz) are still dominated by the single-dipole response of the $n\mhyphen InSb$ particle. This fact can be also seen in the mechanical variables displayed in Figs.~\ref{fig:4_Parallel_conf_varB}b-c, namely, the binding force and the spin torque exerted on particle $\#2$. Nevertheless, these observables display zero values (black lines), which means their behaviour at resonance is more complex than that for the absorption and thus more sensitive. For instance, the binding forces can help detect the second particle and characterize their associated resonances, because $\Delta$ describes the stability of the dimer (around $\Delta=0$) in the momentum space. Specifically, for the $GaSe$ particle, there is a sharp contrast between attraction ($\Delta>0$) and repulsion ($\Delta<0$) between the particles around the characteristic phonon resonance ($\sim 7$ THz). Particle $\#2$ is thus in "stable" (energetic) equilibrium when the illumination frequency is detuned from the phonon resonance. In other terms, more [less] energy in the system implies attractive [repulsive] forces.  
Crucially, the spin torque for the $GaSe$ particle exhibits more atypical behavior, Fig.~\ref{fig:4_Parallel_conf_varB}c. In this case, the values for the torques are almost zero everywhere in the map except for the $GaSe$ phonon resonance's location. Furthermore, the intensity of the spin torque is strongly modified when $B$ increases. This phenomenon can only be explained due to strong interparticle interaction, since particle $\#2$ does not exhibit direct MO properties. Therefore, the spin torque as an observable displays more sensitivity to detect particle $\#2$ than the other variables in the illustrated example.
Noteworthy, the gap between particles may vary on a larger scale to catch the effect of the stationary character of the incident beam. However, these gaps should be as large as a few mm for such effect to be noticed. The absorption grows sharply to reach almost a plateau value that approximately corresponds to the addition of the independent absorptions (particles as isolated, not shown here). The same situation is found for the spin torques, whose growing trend is similar for large gaps. On the contrary, the binding decays rapidly to zero as gap increases. In all the variables, there are oscillations below the order of $0.01 \%$ in the limit of large gaps due to the stationary-wave effect.

\subsection{Dimer under perpendicular configuration}

\begin{figure}
	\centering
		\includegraphics[width=8.5cm,keepaspectratio]{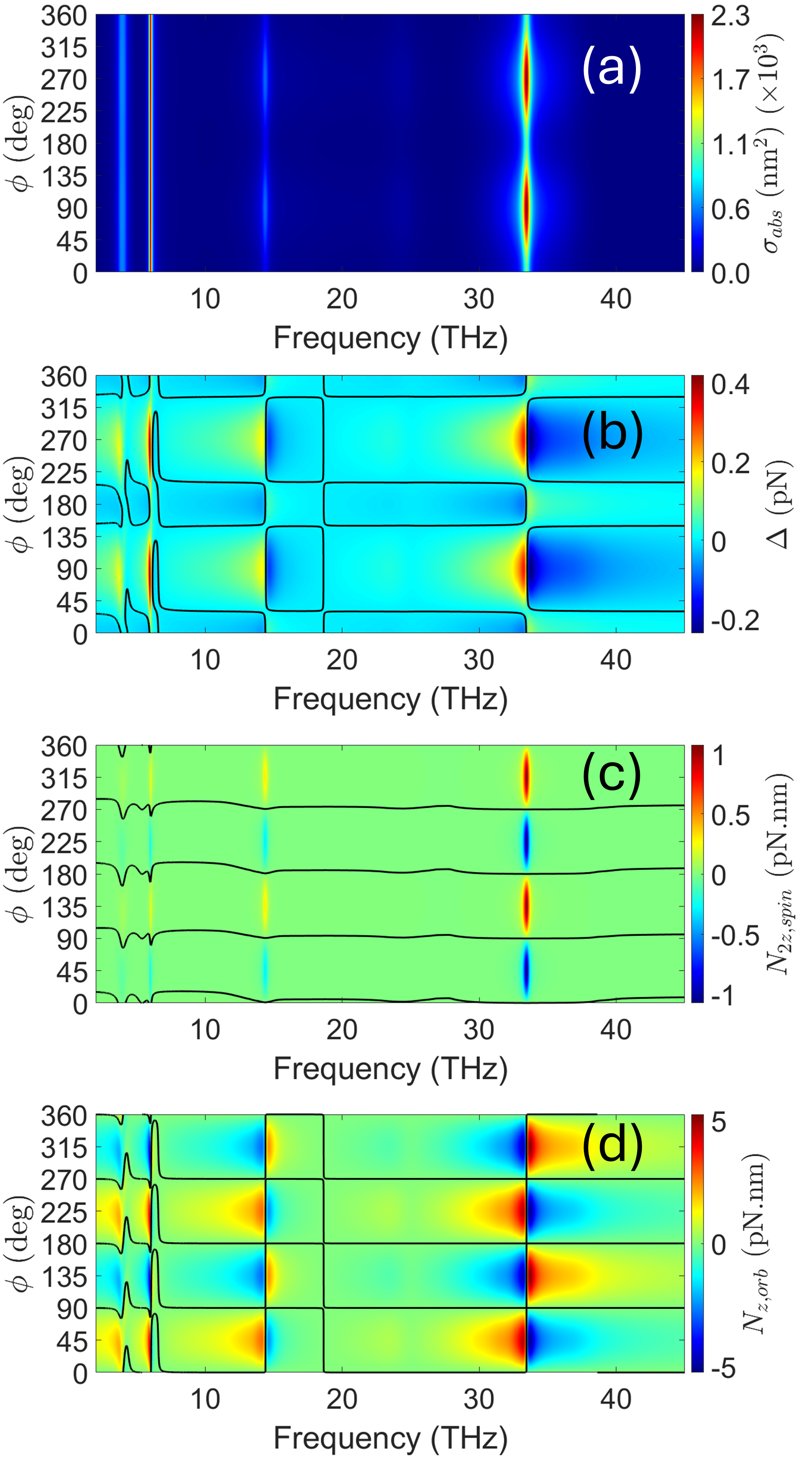}
	\caption{\label{fig:5_Perp_conf_SiO_2} (Color online) Case of the dimer $\left[n\mhyphen InSb, SiO_2\right]$ for perpendicular configuration and $B=0.25$ T. The particles' size are $100$ nm and $50$ nm, respectively, with a $10$ nm gap. (a) Absorption cross section; (b) Binding force, (c) z-Spin torque induced on the second particle. (d) z-Orbital torque induced on the system.}
\end{figure}

Transitioning the dimer to a perpendicular configuration (see Fig.~\ref{fig:1_Config}c) introduces azimuthal dependency. In light of this, the system shows new degrees of freedom that can be described by the radial coordinate $r$ and the azimuthal angle $\phi$, Fig.~\ref{fig:1_Config}c, so that the inter-particle gap is now radial in this frame. The illumination field polarizes the particles according to their angular position. Overall, the illumination of Eq.~\ref{eq-E0} imposes a direction for the electric field which is variable with $z$. Therefore, the beam polarizes the particles according to their $z$-positions. To simplify the interpretation of results, let's assume $z_1 = z_2 = 0$, which imposes a uniform electric field (with a single direction) along the $xy$ plane. With the purpose of exploring the sensing capabilities of the MO heterodimer under this setting, Fig.~\ref{fig:5_Perp_conf_SiO_2} shows the case of the dimer $\left[n\mhyphen InSb, SiO_2\right]$ under a constant field of $B=0.25$ T. Now there are four relevant variables, namely, absorption (Fig.~\ref{fig:5_Perp_conf_SiO_2}a), binding force (Fig.~\ref{fig:5_Perp_conf_SiO_2}b), spin torque exerted on particle $\#2$ (Fig.~\ref{fig:5_Perp_conf_SiO_2}c), and the orbital torque exerted on the whole dimer, Fig.~\ref{fig:5_Perp_conf_SiO_2}d. In particular, the $SiO_2$ particle displays two resonances around $14$ and $33$ THz which correspond to the excitation of phonon polaritons \cite{franta_optical_2016}. Further, these especial resonances are modulated by the dimer angle $\phi$, giving maxima at $90$, $270$ deg (Fig.~\ref{fig:5_Perp_conf_SiO_2}a). This is consistent with the fact that the illumination's electric field at $z=0$ is presently oriented along the $y$-axis. Thus, the dielectric particle makes the dimer notoriously sensitive to the polarization of the illumination field. The binding force displays a similar pattern, with a $45$-deg resonance modulation, Fig.~\ref{fig:5_Perp_conf_SiO_2}b. Note also that the main spots in $\Delta$ coincide with the absorption maxima, located with $90$-deg modulation. Moreover, the binding-force resonances of $SiO_2$ are "stable" in the sense described above for parallel configuration: each spot has a "bimodal" structure due to a striking contrast between negative a positive values near resonance. 

At this point, $\Delta$ shows that also the MO resonances located below $10$ THz are modulated by angle. Therefore, $\Delta$ is more sensitive as observable than the absorption in FF. The absorption does not show such dependency on $\phi$ at low frequencies. Another consequence of this enhanced sensitivity in NFs is the complex pattern of zeros in the force, black lines in Fig.~\ref{fig:5_Perp_conf_SiO_2}b; the curves at low frequencies are strongly dependent on $B$, showing complex shapes. Such effect can also be seen in Figs.~\ref{fig:5_Perp_conf_SiO_2}c-d for the induced torques.

Outstandingly, the torques also expose $45$-deg resonance modulation, Fig.~\ref{fig:5_Perp_conf_SiO_2}c-d. Also, orbital torques show the same kind of bimodal spots as the ones in $\Delta$, Fig.~\ref{fig:5_Perp_conf_SiO_2}d; the colors (signs) of the spots around the resonances also reverse with $90$-deg modulation. For the orbital torques, however, such spots have a $45$-deg phase difference; the "stability" criterion for the orbital torque is different. Another significant remark between both kind of torques is about their kind of spot: $N_{2z}$ has "monomodal" spots instead of the bimodal of the orbital torque, Figs.~\ref{fig:5_Perp_conf_SiO_2}c-d. As a result, the torques and radial forces are synchronized and the dimer must obey to a spin-orbit coupling mediated by light. This coupling allows for a clear method of identifying / detecting particles in the dimer, as the well-resolved, particular spots in Fig.~\ref{fig:5_Perp_conf_SiO_2}c reveal.

\begin{figure}
	\centering
		\includegraphics[width=8.5cm,keepaspectratio]{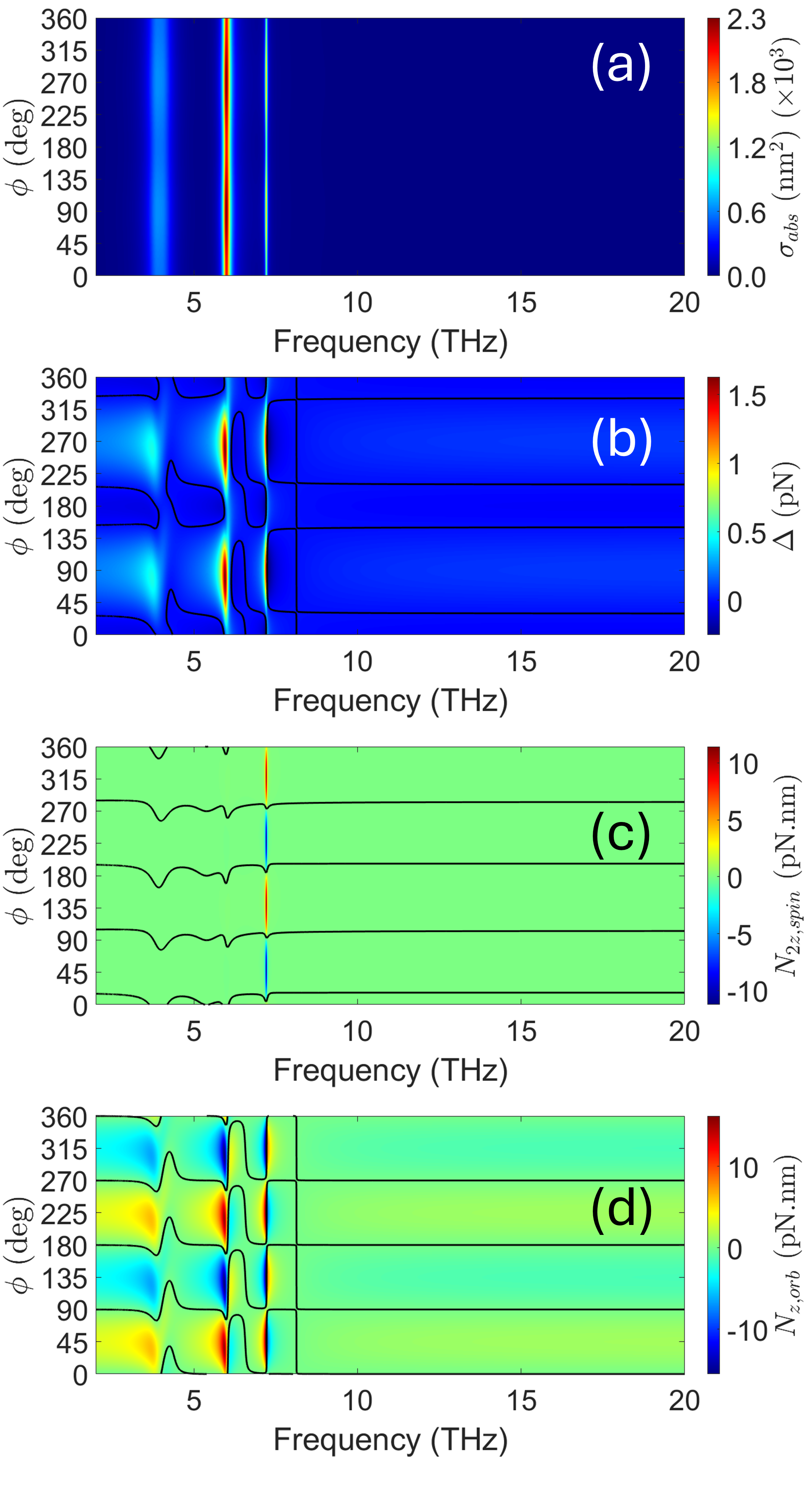}
	\caption{\label{fig:6_Perp_conf_GaSe} (Color online) Case of the dimer $\left[n\mhyphen InSb, GaSe\right]$ for perpendicular configuration and $B=0.25$ T. (a) Absorption cross section; (b) Binding force, (c) Spin torque induced on the second particle. (d) Orbital torque induced on the system.}
\end{figure}

Finally, the response by the dimer $\left[n\mhyphen InSb, GaSe\right]$ under perpendicular configuration is examined for a low field of $B=0.25$ T, Fig.~\ref{fig:6_Perp_conf_GaSe}. In this case, the absorption displays practically the same values as the absorption under parallel configuration, Fig.~\ref{fig:6_Perp_conf_GaSe}a,  compare with Fig.~\ref{fig:4_Parallel_conf_varB}a. Nonetheless, the angular modulation of these resonances is visibly absent. Again, the FF observable lacks relevant information of the dimer, in this case its orientation. In contrast, the binding force exhibits strong modulation with $\phi$ that maximizes attraction at $90$, $270$ deg for all the resonances. Their maxima are an order of magnitude higher than those in Fig.~\ref{fig:4_Parallel_conf_varB}b. In particular, the strongest spots show the bimodal structure (stability) like those shown in Fig.~\ref{fig:5_Perp_conf_SiO_2}b for silica's case. The phonon resonance at $7.2$ THz, corresponding to $GaSe$, appears as strong as the main MO resonance for $n\mhyphen InSb$ at $6$ THz.  The map of Fig.~\ref{fig:6_Perp_conf_GaSe}b becomes naturally more complex than its analogous for Fig.~\ref{fig:4_Parallel_conf_varB}b, showing an intricate resonance pattern of zeros (black lines). Analogous patterns of zeros appear also for the induced torques, Figs.~\ref{fig:6_Perp_conf_GaSe}c-d. In addition, there are stronger and sharper spots than the previous cases analyzed. As an illustration, compare the values obtained for the absorption, binding and torques calculated for $SiO_2$'s case, Fig.~\ref{fig:5_Perp_conf_SiO_2}a-d; although in both systems, $SiO_2$ and $GaSe$, the absorption reaches maxima at a common value $2300$ nm$^2$, the induced mechanical variables result very different. $GaSe$ turns out to be much stronger and more detectable as a particle than $SiO_2$, even though the geometries compared are equal.
The exhaustive study of MO dimers in the THz regime demonstrates that mechanical variables -binding forces, spin torques, and orbital torques- provide a superior and more nuanced understanding of light-matter interactions at the nanoscale \cite{salikhov_spin-orbit_2025}. Unlike conventional FF spectroscopy, which often masks the contribution of individual sub-units within a cluster, NF mechanical probing acts as a high-resolution, dynamic diagnostic tool capable of identifying particles based on their specific resonant behaviors and crystallographic orientations.  
 
Overall, the use of materials like $n\mhyphen InSb$, which exhibits a significant MO response at low fields, and materials like $GaSe$, which provides sharp, anisotropic phonon signatures, allows for the creation of tunable, non-reciprocal systems. These systems can be effectively calibrated and monitored even when immersed in complex, viscoelastic biological mimics such as PVA hydrogels or paraffin wax. The sensitivity of these mechanical variables can be summarized by their ability to resolve the "hidden" components of a dimer: \\
1.	Parallel Binding Forces: Act as proximity sensors, identifying particle $\# 2$'s presence and resonant stability.  \\ 
2.	Spin Torques: Function as material identifiers, showing distinct sign changes and zeros that correlate with internal phonon and plasmon modes. \\  
3.	Orbital Torques: Provide orientation sensitivity, revealing the azimuthal alignment of a dimer in a structured field. These are "readers" of the field's polarization. \\  

In essence, the proposed mechanical variables are in all cases studied more accurate and useful than the FF's absorption as to serve for detection and sensitive reading of nearby particles. The examples shown support the fact that simple MO dimer systems could help detect [/work as] THz sources even in complex environments as biological media. The presented results pave the way for assessing THz methods of in-vivo theragnostics, THz communication, and design of nanomachines at low energies \cite{stanton_magnetotactic_2017}. As a final thought, more complex systems than dimers could also be useful for detection and sensitivity in the embedding medium, such as multimers made of particles, pillars, or even waveguides and other MO couplers. These, however, would requiere further investigation provided they are expected to offer more degrees of freedom in terms of dynamics as well as more challenges to calibrate and control them. 

\section{Conclusions}

A new concept of near-field sensing has been theoretically demonstrated through the study of simple systems consisting of optically-coupled nanoparticles. This framework allows for the construction of particle sensors and nanoscale calibration methods by utilizing the interaction between magneto-optical (MO) nanoparticles and nearby particles of any constitutive medium. Such systems are shown to be effective even when immersed in complex environments, such as biological phantoms, where far-field diffraction limits typically hinder characterization.

The proposed mechanical variables represent a radical advancement in the field of THz sensing by moving beyond the limitations inherent to the far-field. By selecting dynamic observables with a direct relation to the local near-fields, the environment of nanoparticle clusters can be explored with a degree of precision and accuracy previously unattainable. This approach leads directly toward the development of next-generation tools for nanomedicine, improved communication systems, and highly efficient nanomachines. While challenges remain in system miniaturization and clinical integration, the development of fiber-coupled nano-optomechanical sensors and portable THz metasurface platforms will be critical for transitioning these principles to point-of-care diagnostics.

Beyond the calibration potential of isolated particles, the analysis of heterodimer systems highlights the emergence of unique mechanical fingerprints that are absent in far-field observables. Notably, for the $\left[n\mhyphen InSb, GaSe\right]$ dimer, the spin torque on the non-MO GaSe particle is driven entirely by strong near-field interaction with the neighboring MO particle. This coupling allows for the indirect magnetic control of the GaSe particle's mechanical state via the external static magnetic field, manifesting as a highly localized spectral signature at the GaSe phonon resonance. Furthermore, the sharp contrast between attractive and repulsive binding forces provides a 'stability' criterion for characterizing these associated resonances and defining stable equilibrium points for particle trapping. These findings underscore the superior sensitivity of near-field mechanical variables for identifying 'hidden' particles and resolving inter-particle coupling.
\begin{acknowledgments}
	
	R.M.A-E. would like to thank D.T. for his continuous support and encouragement to complete this research. Also GROC, UJI, UNCPBA, CICPBA, and CONICET are sincerely acknowledged for providing the necessary office and time resources that significantly contributed to the completion of this research. 
		
	This project has received funding from the European Union’s Horizon Europe research and innovation programme under Grant Agreement No 101046489 (EIC Pathfinder project DYNAMO).
\end{acknowledgments}

\appendix 

\section{Lorentz model for Gallium Selenide}

Following previous reports for this material, the dielectric function tensor components used in Eq.~\ref{eq-epr-birrefringence} were implemented using a Lorentzian oscillator model:
\begin{align}
	\epsilon_\perp &= \epsilon_{\infty, \perp} + \frac{S_\perp \omega^2_{TO, \perp}}{\omega^2_{TO, \perp}- \omega^2 - \mathrm{i} \gamma_\perp \omega}, \\
	\epsilon_\parallel &= \epsilon_{\infty, \parallel} + \frac{S_\parallel \omega^2_{TO, \parallel}}{\omega^2_{TO, \parallel} - \omega^2 - \mathrm{i} \gamma_\parallel \omega}	
\end{align}
where the parameters for the ordinary ($\perp$) and extraordinary ($\parallel$) components are:
$\epsilon_{\infty, \perp} = 7.44$ and   $\epsilon_{\infty, \parallel} = 5.76$, for the high-frequency dielectric constants; $\omega_{TO, \perp} = 4.0310 \times 10^{13}$ rad/s and $\omega_{TO, \parallel} = 4.4454 \times 10^{13}$ rad/s for the transverse optical phonon frequencies; $S_\perp = 3.16$ and $S_\parallel = 2.50$ for the oscillator strengths; and $\gamma_\perp = 3.7673 \times 10^{11}$ rad/s and $\gamma_\parallel = 5.6510 \times 10^{11}$ rad/s for the damping constants. These values align perfectly with well-established experimental data for single crystals in this material \cite{chen_generation_2006,chen_optical_2009}.


%

\end{document}